\documentclass[iop]{emulateapj}

\def\microas{\mu{\rm as}}
\def\muas{\mu{\rm as}}

\begin{document}

\title{The PHASES Differential Astrometry Data Archive.  I.  Measurements and Description}

\author{Matthew W.~Muterspaugh\altaffilmark{1, 2}, 
Benjamin F.~Lane\altaffilmark{3}, 
S.~R.~Kulkarni\altaffilmark{4}, 
Maciej Konacki\altaffilmark{5, 6}, 
Bernard F.~Burke\altaffilmark{7}, 
M.~M.~Colavita\altaffilmark{8}, 
M.~Shao\altaffilmark{8}, Sloane J.~Wiktorowicz\altaffilmark{9}, 
J.~O'Connell\altaffilmark{1}}
\altaffiltext{1}{Department of Mathematics and Physics, College of Arts and 
Sciences, Tennessee State University, Boswell Science Hall, Nashville, TN 
37209 }
\altaffiltext{2}{Center of Excellence in Information Systems, Tennessee 
State University, 3500 John A. Merritt Boulevard, Box No.~9501, Nashville, TN 
37209-1561}
\altaffiltext{3}{Draper Laboratory,  555 Technology Square, Cambridge, MA 
02139-3563}
\altaffiltext{4}{Division of Physics, Mathematics and Astronomy, 105-24, 
California Institute of Technology, Pasadena, CA 91125}
\altaffiltext{5}{Nicolaus Copernicus Astronomical Center, Polish Academy of 
Sciences, Rabianska 8, 87-100 Torun, Poland}
\altaffiltext{6}{Astronomical Observatory, Adam Mickiewicz University, 
ul.~Sloneczna 36, 60-286 Poznan, Poland}
\altaffiltext{7}{MIT Kavli Institute for Astrophysics and Space Research, 
MIT Department of Physics, 70 Vassar Street, Cambridge, MA 02139}
\altaffiltext{8}{Jet Propulsion Laboratory, California Institute of 
Technology, 4800 Oak Grove Drive, Pasadena, CA 91109}
\altaffiltext{9}{Department of Astronomy, University of California, Mail Code 
3411, Berkeley, CA 94720, USA}

\email{matthew1@coe.tsuniv.edu, blane@draper.com, maciej@ncac.torun.pl}

\begin{abstract}
The Palomar High-precision Astrometric Search for Exoplanet Systems (PHASES) 
monitored 51 sub-arcsecond binary systems to determine precision binary 
orbits, study the geometries of triple and quadruple star systems, and 
discover previously unknown faint astrometric companions as small as giant 
planets.  PHASES measurements made with the Palomar Testbed Interferometer 
(PTI) from 2002 until PTI ceased normal operations in late 2008 are 
presented.  Infrared differential photometry of several PHASES targets were 
measured with Keck Adaptive Optics and are presented.
\end{abstract}


\section{Introduction}

A technique has been developed to obtain high precision (35 $\microas$) 
astrometry of close stellar pairs 
\citep[separation less than 1 arcsec;][]{LaneMute2004a} using 
long-baseline infrared interferometry at the Palomar Testbed 
Interferometer \citep[PTI;][]{col99}.  This technique was applied to 51 binary 
systems as the Palomar High-precision Astrometric Search for Exoplanet Systems 
(PHASES) program during 2002-2008.  PHASES science results included precision 
binary orbits and component masses, studies of the geometries of and 
physical properties of stars 
in triple and quadruple star systems, and limits on the 
presence of giant planet companions to the binaries.

This paper is the first in a series analyzing the final results of the PHASES 
project as of its completion in late 2008.  This paper describes the 
observing method, sources of measurement uncertainties, limits of observing 
precisions, derives empirical scaling rules to account for noise sources 
beyond those predicted by the standard reduction algorithms, and presents the 
full catalog of astrometric measurements from PHASES.  The second paper 
combines PHASES astrometry, astrometric measurements made by other methods, 
and radial velocity observations (where available) to determine orbital 
solutions to several binaries' Keplerian motions, determining physical 
properties such as component masses and system distance when possible 
\citep{Mute2010B}.  The third paper presents limits on the existence of 
substellar tertiary companions orbiting either the primary or secondary stars 
in those systems that are found to be consistent with being simple binaries 
\citep{Mute2010C}.  Paper four presents orbital solutions to a known triple 
star system (63 Gem A $=$ HD 58728) and a newly discovered triple system 
(HR 2896 $=$ HD 60318) \citep{Mute2010D}.  Finally, paper five 
presents candidate substellar companions to PHASES binaries as detected by 
astrometry \citep{Mute2010E}.

Astrometric measurements were made at PTI, which was 
located on Palomar Mountain near San Diego, CA. It was developed by the Jet 
Propulsion Laboratory, California Institute of Technology for NASA, as a 
testbed for interferometric techniques applicable to the Keck Interferometer 
and other missions such as the Space Interferometry Mission (SIM).  It 
operated in the J ($1.2 \mu{\rm m}$), H ($1.6 \mu{\rm m}$), and K 
($2.2 \mu{\rm m}$) bands, and combined starlight from two out of three 
available 40-cm apertures.  The apertures formed a triangle with one 110 and 
two 87 meter baselines.  PHASES observations began in 2002 and 
continued through 2008 November when PTI ceased routine operations.

\section{Observations and Data Processing}

The initial PHASES observing method and data processing algorithm were 
presented by \cite{LaneMute2004a}.  Incremental improvements to these 
procedures were updated in papers from the PHASES science program.  The final 
observing procedure and data processing algorithm is presented in complete form 
here.  All astrometry measurements were reprocessed using the final algorithm 
presented here.  Measurements taken with different instrumental configurations 
than the standard one presented here (for example, those lacking longitudinal 
dispersion compensation) are noted in Table \ref{phasesAstrometryData}.

\subsection{Astrometric Observation Method}

\subsubsection{Optical Interferometers}

In an optical interferometer, light is collected at two or more
apertures and brought to a central location where the beams are
combined and a fringe pattern produced on a detector 
(at PTI, the detectors were NICMOS and HAWAII 
infrared arrays, of which only a few 
pixels were used).  For a broadband source of central 
wavelength $\lambda$ and optical bandwidth $\Delta\lambda$ 
(for PTI $\Delta\lambda = 0.4 \mu$m), 
the fringe pattern is limited in extent and
appears only when the optical paths through the arms of the
interferometer are equalized to within a coherence length ($\Lambda =
\lambda^2/\Delta\lambda$). For a two-aperture interferometer,
neglecting dispersion, the intensity measured at one of the combined
beams is given by
\begin{equation}\label{double_fringe}
I(x) = I_0 \left ( 1 + V \frac{\sin\left(\pi x/ \Lambda\right)}
{\pi x/ \Lambda} \sin \left(2\pi x/\lambda \right ) \right )
\end{equation}
\noindent where $V$ is the fringe contrast or ``visibility'', which
can be related to the morphology of the source, 
and $x$ is the optical path difference between arms of the 
interferometer.  More detailed analysis 
of the operation of optical interferometers can be found in {\it Principles of 
Long Baseline Stellar Interferometry} \citep{Lawson2000}.

\subsubsection{Interferometric Astrometry}

The location of the resulting interference fringes is 
related to the position of the target star and the observing geometry
via
\begin{equation}\label{delayEquation}
d = \overrightarrow{B} \cdot \overrightarrow{S} + \delta_a\left(\overrightarrow{S}, t\right) + c 
\end{equation}
\noindent where $d$ is the optical path length one must introduce
between the two arms of the interferometer to find fringes. This
quantity is often called the ``delay.'' $\overrightarrow{B}$ is the 
baseline---the vector 
connecting the two apertures. $\overrightarrow{S}$ is the unit vector
in the source direction, and $c$ is a constant additional scalar delay
introduced by the instrument.  
The term $\delta_a\left(\overrightarrow{S}, t\right)$ 
is related to the differential amount of path introduced by the atmosphere 
over each telescope due to variations in refractive index.
For a 100-m baseline interferometer an astrometric precision of 10 $\mu$as
corresponds to knowing $d$ to 5 nm; while difficult, this is achievable for 
all terms except that related to the atmospheric delay.  
Atmospheric turbulence, which changes over
distances of tens of centimeters and on millisecond timescales, forces
one to use very short exposures (to maintain fringe contrast) and
limits the sensitivity of the instrument.  It also severely limits
the astrometric accuracy of a simple interferometer, at least over 
large sky-angles.

However, in narrow-angle astrometry one is concerned with a close pair of 
stars, and the observable is a differential astrometric measurement, i.e.~one 
is interested in knowing the angle between the two stars 
($\overrightarrow{\Delta_s} = \overrightarrow{s_2} - \overrightarrow{s_1}$).  
The atmospheric turbulence is correlated over small angles.  If the 
measurements of the two stars are simultaneous, or nearly so, the atmospheric 
term cancels out.  Hence, it is still possible to obtain high precision
``narrow-angle'' astrometry.

\subsubsection{Sub-Arcsecond Differential Astrometry}\label{sec::lightpath}

\begin{figure*}[!ht]
\plotone{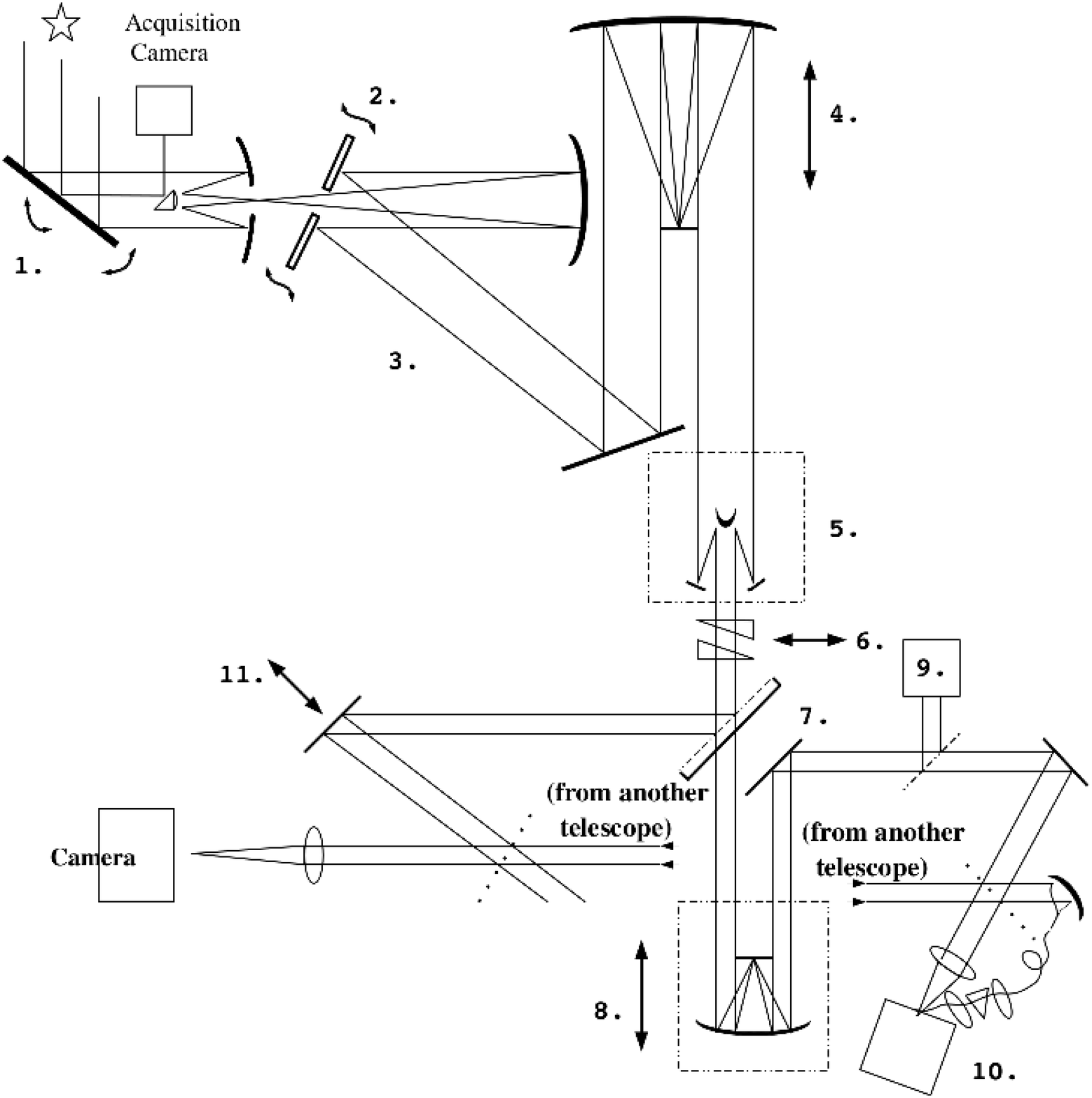}
\caption[PHASES Instrument Configuration and Light Path]
{ \label{fig:phasesLightPath}
Configuration and light path used at PTI for the PHASES experiment is 
shown in schematic form; the corresponding description is described in Section 
\ref{sec::lightpath}.  Note that the path for only one telescope is shown; 
light from the second telescope travels a similar path.
}
\end{figure*}

As illustrated in Figure \ref{fig:phasesLightPath}, 
the light path of the PHASES astrometry program was as follows:
\begin{enumerate}
\item Light was collected by two of the three siderostats at PTI.  The 
siderostats had 
diameter 50 cm and fed 40 cm beam compressing telescopes.  These collimated 
the 40 cm input into a 7.5 cm beam.
\item At the siderostat enclosure, a fast steering mirror (FSM) provided 
tip-tilt 
correction for low-order adaptive optics improvement.  The feedback sensor for 
the tip-tilt system was located in the beam combining facility (see step 
(9)), to include both atmospheric and instrumental sources of tip-tilt 
variations.
\item The collimated beams propagated through pipes to the central beam 
combining building.  For the north and south siderostats, the pipes were held 
at vacuum; for the west siderostat, the pipe was filled with air.
\item Each collimated beam was directed to a long delay line (DL) which 
tracks the sidereal delay rate and receives feedback from the fringe tracking 
beam combiner for removing some of the atmospheric turbulent delay 
variations.  There were roughly $\pm 38$ m of 
optical delay available in the long DLs.
\item The collimated beams were recompressed to 2.5 cm collimated beams.
\item Each beam passed through a pair of matched prisms.  In one arm, one 
prism was on a linear stage to vary the total glass thickness the starlight 
passed through.  The positions of these prisms were calibrated using fringes 
from an internal light source to minimize the combined longitudinal dispersion 
of air path and glass.  This calibration also determined the glass-to-air 
dispersion correction rate.  The linear stage was operated in open-loop mode 
to minimize longitudinal dispersion even as the air path was varied by the 
long DLs.  Some early measurements were made before this dispersion 
compensator was available, as marked in Table \ref{phasesAstrometryData}.
\item The 2.5 cm collimated beams were split $\sim 70/30$ by plate beam 
splitters.  The reflected $\sim 30$\% of the light was directed to the 
``secondary'' beam combiner which made the astrometric science measurement.  
Red He-Ne LASER metrology signals were injected at these beam splitters and 
propagated through both beam combiners before being extracted just before the 
infrared array detectors to monitor path variations.
\item The $\sim 70$\% transmitted light from each beam was directed through a 
short DL.  These were smaller versions of the long DLs with a few ten's of 
centimeters of travel.  The short DLs allowed one to set the delay 
offset between the fringe-tracking and the science beam combiners such that 
both are near zero optical delay simultaneously.  It also introduced the 50-100 
Hz sawtooth modulation required for sampling fringes in the fringe tracker, 
without being in the science camera's optical path.
\item After the short DL, the $\sim 70$\% of the original light to be 
used for fringe tracking was directed to the ``primary'' beam combiner.  I-band 
light was extracted using a dichroic beamsplitter and focused onto quad-cell 
APD's.  The signal was used as feedback to the fast tip-tilt mirrors in step 
2.
\item The longer-wavelength light was combined using a plate beam splitter.  
Half the light was focused on a single pixel of the 
NICMOS infrared array for high-signal-to-noise ratio (S/N) 
phase tracking.  The other half passed through a single-mode fiber to improve 
wavefront quality (and system visibility), then through a low-resolution prism 
to disperse the light onto $\sim 5$ pixels for slower group-delay tracking.  
Phase and group-delay signals were fed to the long DLs to reduce jitter 
from atmospheric piston motions.  This beam combiner was the same as used for 
PTI's standard visibility mode observations.
\item The $\sim 30$\% of the light redirected to the ``secondary'' table was 
used for the astrometric science measurement.  The secondary beam combiner was 
identical to the first with the following exceptions.
  \begin{enumerate}
  \item A piezo-driven scanning mirror added up to $\pm 300$ $\mu$m of optical 
        delay to allow scanning within the arcsecond field of view.  The scan 
        was close-loop controlled using the LASER metrology signal.
  \item No single-mode fiber was used for spatial filtering.
  \item A HAWAII array was used instead of NICMOS.
  \end{enumerate}
\end{enumerate}

The entire optical system was realigned at the beginning of the night and 
usually updated in the middle of the night, once per night.  Alignment drifts 
from one night to the next were small.  Standard detector calibrations (gain, 
bias and background) were acquired at the beginning of the night for both 
infrared array detectors; background levels were re-measured with each star 
acquisition.

The operating sequence began by acquiring the stars and locking star trackers 
for tip-tilt corrections on both telescopes.  The 50-100 Hz fringe-tracking 
modulation was applied to the long DLs, which were moved until fringes 
were found with the ``primary'' beam combiner.  Once fringe lock was made, the 
long and short DLs were moved simultaneously to maintain primary 
fringes until fringes were detected on the ``secondary'' beam combiner.  Once 
fringes were found simultaneously on both beam combiners, (1) the zero-point 
offset for the short DLs was recorded, (2) the 50-100 Hz fringe 
tracking modulation was moved from the long to short DLs so that only 
the ``primary'' beam combiner was affected by the modulation, (3) the 
``secondary'' detector's read-out pattern was adjusted from one optimized for 
fringe tracking to a faster and more evenly spaced one that was 
better for scanning 
between fringe packets, and (4) the scanning mirror in the ``secondary'' beam 
combiner initiated its $\sim$Hz, $\sim 100$ $\mu$m triangle modulation.  The 
speed and amplitude of the triangle pattern wer user-settable according to the 
predicted projected binary separation.  Conservative values were used with 
larger amplitudes than predicted to ensure good coverage of the fringe packets.
Observing a binary when its baseline projected separation 
($\overrightarrow{B} \cdot \overrightarrow{\Delta S}$) was of order the 
interferometric coherence length 
($\Lambda = \lambda^2/\Delta\lambda \approx 20 {\rm \mu m}$) or less 
was avoided due to potential biases associated with an imperfect 
template fringe packet.

This setup resulted in a special astrometry mode designed to work on pairs of 
stars separated by no more than $\sim 1$ arcsec, the diffraction limit of 
the 40 cm apertures at $\lambda = 2.2 \, {\rm \mu m}$.  In this mode, the 
small separation of the binary resulted in both binary components being in the 
field of view of a single interferometric beam combiner.  The fringe positions 
were measured by modulating the instrumental delay with an amplitude large
enough to record both fringe packets.  This eliminated the need for a
complex internal metrology system to measure the entire optical path
of the interferometer, and dramatically reduced the effect of
systematic error sources such as uncertainty in the baseline vector
(error sources which scale with the binary separation).  The same instrumental 
configuration could be used for double Fourier spectroscopy.

However, since the fringe position measurement of the two stars was no longer 
truly simultaneous it was possible for the atmosphere to introduce path-length 
changes (and hence positional error) in the time between measurements of the 
separate fringes.  To reduce this effect a fraction of the incoming starlight 
was redirected to a separate beam-combiner, as described above.  
This beam-combiner was used in a
``fringe-tracking'' mode \citep{ss80,col99} where it rapidly (10-20 ms) 
measured the phase of one of the starlight fringes and adjusted the internal 
delay to keep that phase constant.  The fringe tracking data were used both in 
real-time (operating in a feed-back servo, after which a small---but 
measurable---residual phase error remained) and in post-processing (the 
measured residual error was applied to the data as a feed-forward servo).  This 
technique---known as phase referencing---had the effect of stabilizing the 
fringe measured by the astrometric beam-combiner.  For this observing mode, 
LASER metrology was only required between the two beam combiners through 
the location of the light split (which occurred after the optical delay has been 
introduced), rather than throughout the entire array.  This greatly reduced 
system complexity and increased the percentage of time on-sky.  Extra efforts 
in system reliability and automation allowed most PHASES measurements to be 
acquired by a single night assistant with high ($> 90$\%, not counting weather) 
on-sky efficiency.

In making an astrometric measurement the optical delay was modulated in a 
triangle-wave pattern around the stabilized fringe position, while measuring 
the intensity of the combined starlight beams.  The range of the delay sweep 
was set to include both fringe packets; typically this required a scan 
amplitude on the order of $150~\mu$m.  Typically one such ``scan'' was obtained 
every second, consisting of up to 1000 intensity samples (the scan rate was 
limited by the source brightness and the requirement that $>2$ samples are 
made per fringe modulation period).  A double fringe packet based on 
Equation \ref{double_fringe} was then fit to the data, and the differential 
optical path between fringe packets was measured.

\subsection{Astrometric Data Reduction Algorithm} \label{sec::reduction}

\begin{figure*}[!ht]
\plotone{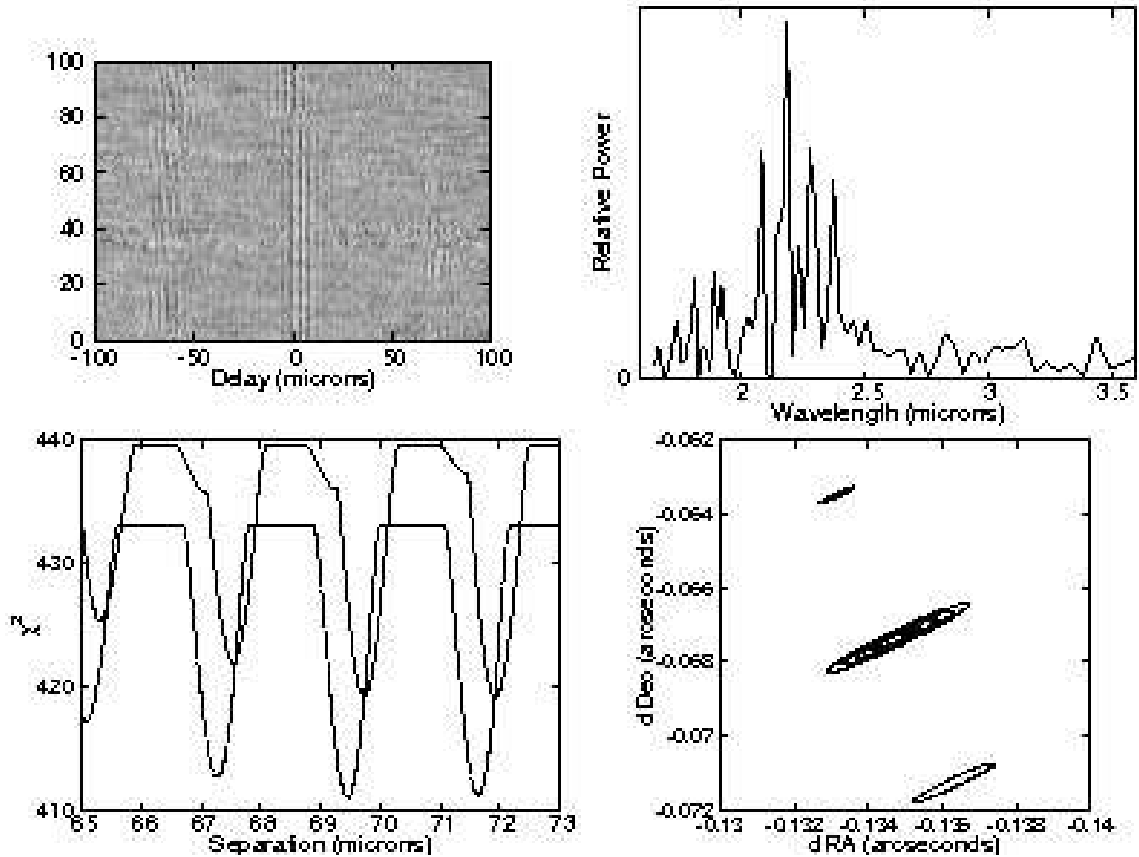}
\caption[Data Processing] 
{ \label{fig::dataProcessing}
Several steps of the data processing pipeline are shown for an observation of 
HD 13872 (21 Ari) from MJD 54385.33270.  Upper left:  
raw interferograms show two distinct fringe packets, one from each star 
in the over-resolved binary.  One star was found 
near zero delay, the other at 
$\sim \pm 70$ ${\rm \mu m}$.  The flipping from 
side-to-side was a result of the 
fringe tracker first tracking on one star for some scans, then the other.  
Upper right:  the periodogram of one scan, used 
to evaluate the fringe S/N within the optical passband used (K-band, 2.0-2.4 
${\rm \mu m}$).  The power in the passband is 
not flat, but rather oscillates, a 
result of the object being a binary star (visibility changes with wavelength).
Lower left:  the likelihood metric $\chi^2$ as a function of delay separation 
of the binaries, for two scans.  Note there are multiple minima separated by 
the fringe modulation period, and the noise is large enough to prevent high 
confidence identification of the correct local minimum.  Lower right:  Tthe 
coadded $\chi^2$ surface in differential R.A./decl., 
with contours corresponding 
to 1-, 2-, 3-, and 4-$\sigma$ confidence regions.  Note that in this case 
sidelobes appeared at the 4-$\sigma$ confidence level, so the measurement was 
rejected as ambiguous.
}
\end{figure*}

The intensity versus delay position measurements produced by the
interferometer were processed into astrometric measurements as
follows.
\begin{enumerate}
\item Detector calibrations (gain, bias, and background) were applied to the 
intensity measurements.
\item The residual phase errors from the primary fringe tracker were converted 
to delay and applied to the data. Note that while the intensity measurements 
were spaced regularly in time, and the delay scanned linearly in time, the 
variable amount of delay correction applied from the fringe tracker resulted 
in the intensity measurements being unevenly spaced in delay. This somewhat 
complicated the downstream processing, in that FFT-based algorithms could not 
be used.
\item The data were broken up into ``scans'' either when the delay sweep 
changed direction or when the fringe tracker lost lock.  Only scans for which 
at least 90\% of the scan was continuously recorded (without fringe lock loss) 
were used in processing.
\item For each scan, a power spectrum was calculated using a Lomb-Scargle 
algorithm \citep{scargle82,press92}.  This spectrum provided an S/N estimate 
based on the ratio of the power in and out of the instrument bandpass.  Only 
the scans with an S/N greater than unity were kept. 
\item For a range of 
values of differential optical delays between fringe packets, a 
model of a double-fringe packet was calculated and compared to the observed 
scan to derive a $\chi^2$ value versus differential delay.
\item A two-dimensional grid in differential R.A.~and decl.~over which to 
search was constructed (in ICRS 2000.0 coordinates).  For each point in the 
search grid the expected differential delay was calculated based on the 
interferometer location, baseline geometry, and time of observation for each 
scan.  These conversions were simplified using the routines from  the Naval 
Observatory Vector Astrometry Subroutines C Language Version 2.0 (NOVAS-C; 
see \cite{novas}).  The value of $\chi^2$ for the grid point's differential 
delay as determined by the previous step was co-added to an 
R.A./decl.~$\chi^2$ grid.  
\item After co-adding the R.A./decl~$\chi^2$ over all the scans in one night, 
all resulting values of $\chi^2$ within $4\sigma$ of the minimum point were 
used to fit a two-dimensional 
quadratic function to interpolate the location of the best 
minimum $\chi^2$ value as the final astrometric solution, and the widths of 
the quadratics determined the formal uncertainties as discussed in detail 
below.  The final product was a measurement of the apparent vector between the 
stars and associated uncertainty ellipse.  Because the data were obtained with 
a single-baseline instrument, the resulting error contours are very 
elliptical, with aspect ratios at times $\ge 10$.
\end{enumerate}
Sample illustrations of several of these stages are plotted in Figure 
\ref{fig::dataProcessing}.

In previous data reductions, the scans were optionally digitally bandpass 
filtered before being processed for astrometry.  This resulted in astrometric 
measurements that differed from those derived without filtering by much less 
than the measurement uncertainties.  In this final analysis, the scans were 
not bandpass filtered.

\subsubsection{Probability Distribution Function Sidelobes}

One potential complication with fitting a fringe to the data was that there 
were 
many local minima spaced at multiples of the operating wavelength. If one were 
to fit a fringe model to each scan separately and average (or fit an 
astrometric model to) the resulting delays, one would be severely limited by 
this fringe ambiguity (for a 110-m baseline interferometer operating at 
$2.2 \mu$m, the resulting positional ambiguity is $\sim 4.1$ 
mas). However, by using the $\chi^2$-surface approach, and 
co-adding the probabilities associated with all possible delays for each scan, 
the ambiguity disappeared.  This was due to two things, the first being that 
co-adding simply improved the S/N.  Second, since the 
observations usually lasted for an hour or even longer, the associated baseline 
change due to Earth rotation also had the effect of ``smearing'' out all but 
the true global minimum.  The final $\chi^2$-surfaces did have dips separated 
by $\sim 4.1$ mas from the true location, but any data sets for 
which these showed up at the $4\sigma$ level were rejected.  The $4\sigma$ 
region is that for which the properly normalized $\chi^2$ function has a 
value less than the number of degrees of freedom plus 19.33, the latter being 
the value appropriate for two-dimensional $\chi^2$ distributions.  The final 
astrometry measurement and related uncertainties were derived by fitting only 
the $4\sigma$ region of the surface.

\subsubsection{Residual Unmonitored Phase Noise}

Unmonitored system phase noise can affect the $\chi^2$ surface in two ways.  
First, components of the phase noise that operated at frequencies faster than 
the scan rate caused the two fringe packets to be smeared an extra amount, and 
to first order this appeared as extra noise in the intensity measurements.  
This affected the width of the $\chi^2$ fit for each individual scan (which is 
designated as $\sigma_m$, the ``measurement'' noise), and thus appeared 
directly in the co-added $\chi^2$ contour.

If instead the instrumental noise was much slower than an individual scan, it 
was essentially ``frozen into'' the scan---for the duration of that scan, the 
stars really did appear to have a different separation than their true 
separation.  The $\chi^2$ surface for the fit to an individual scan takes the 
form 
\begin{equation}
f(d-d_j) = \frac{\left(d-d_j\right)^2}{\sigma_m^2} + n
\end{equation}
\noindent where $d_j$ is the value of the star separation that minimizes 
$f=\chi^2$, and $n$ is the number of degrees of freedom of the fit (typical 
values for $n$ are 400--1000; for this derivation, it suffices to assume a 
one-dimensional $\chi^2$ surface as it has no curvature in the direction 
perpendicular to the sky-projected baseline---only Earth-rotation synthesis 
lifts this degeneracy).  The low-frequency components of the phase noise cause 
$d_j$ to vary from $d_o$, the true star separation, by more than one expects 
from measurement noise alone.  By taking many such scans, one can determine 
this instrumental scatter (which is designated as $\sigma_i$, the 
``instrument'' noise for an individual scan) and add (in quadrature) the 
instrumental noise to the measurement noise as 
\begin{equation}
F(d-d_o) = \frac{\left(d-d_o\right)^2}{\left(\sigma_m^2 + \sigma_i^2\right)/N} + nN
\end{equation}
\noindent where $N$ is the number of scans ($N$ was typically hundreds to 
thousands).

Consider a function $f(d-d_{o_j})$ with position of the minimum at $d_{o_j}$; 
this centroid position is distributed with probability
\begin{equation}
P\left(d_{o_j}\right) = \frac{e^{-\left(d_o - d_{o_j}\right)^2/2\sigma_i^2}}{\sqrt{2\pi}\sigma_i}.
\end{equation}
\noindent One may naively hope that summing several instances of this function 
with variable $d_{o_j}$ together would properly add the instrumental and 
measurement noises in quadrature.  However, the summation results in 
\begin{eqnarray}
\sum_{j=0}^N f\left(d - d_{o_j}\right) & = &
N \int_{-\infty}^{\infty} f\left(d-x\right) P\left(x\right) {\rm d} x \\
& = & \frac{\left(d-d_o\right)^2}{\left(\sigma_m^2\right)/N} + nN + N\frac{\sigma_i^2}{\sigma_m^2} \\
& \ne & \frac{\left(d-d_o\right)^2}{\left(\sigma_m^2 + \sigma_i^2\right)/N} + nN.
\end{eqnarray}
\noindent Even if one renormalizes so that the additive term equals $nN$ 
(i.e.~multiply by $n/(n + \sigma_i^2/\sigma_m^2)$), this is still:
\begin{equation}
\sum_{j=0}^N f\left(d - d_{o_j}\right) = 
\frac{\left(d-d_o\right)^2}{\left(\sigma_m^2 + \sigma_i^2/n\right)/N} + nN.
\end{equation}
\noindent Note the extra factor of $n$ dividing $\sigma_i^2$; this effectively 
underestimates the scan-to-scan instrumental noise by a very large 
amount---roughly $20\times$ for typical PHASES data.

Instead, the appropriate way to determine the scan-to-scan fit is by noticing 
that the minimum value of the co-added $\chi^2$ surface is greater than the 
total number of degrees of freedom $nN$ by the amount:
\begin{equation}
N\frac{\sigma_i^2}{\sigma_m^2}.
\end{equation}
\noindent The quantity $\sigma_m$ was measured directly from the shape of the 
surface, which remained 
unchanged, and the number of scans $N$ was known.  Thus, one 
could derive $\sigma_i$ and apply it to the formal uncertainties.  For all 
observations the average value of $\sigma_i^2/\sigma_m^2$ 
was 1.29; values ranged from 0.0042 (for bright sources and good weather 
conditions) to 7.2.

Phase-referencing was used to decrease the amount of unmonitored phase noise 
during narrow-angle astrometry observations (see Section \ref{obsNoise}), but 
some residual phase noise remained, so the correction outlined here had to be 
applied to the astrometric data.  Synthetic data were constructed both 
with and without unmonitored phase noise of the actual spectrum observed, and 
the data reduction algorithm determined measurement uncertainties consistent 
with the actual scatters in the measurements between multiple synthetic data 
sets.  Without the additional phase-noise correction outlined here, the formal 
uncertainties significantly underestimated the scatter in the results.

\section{Expected Performance}\label{sec::expectedperf}

The expected astrometric performance of the PHASES observing mode was 
determined by several factors contributing measurement uncertainties and 
biases.  These can be subdivided into three broad categories:
\begin{enumerate}
\item observational noise terms, which are fundamental to atmospheric 
turbulence and finite source brightness; 
\item instrumental noise terms, which result from the design of the 
interferometer and the method in which the measurements were obtained; 
\item astrophysical noise terms, which result from the astrometric stability 
of the stars themselves.
\end{enumerate}
The size of each noise source is summarized in Table 
\ref{noiseTermSummaryTable}.  The details of a few of these terms were 
described by \cite{LaneMute2004a}; here, a more complete summary of the PHASES 
error budget is presented.

\begin{deluxetable}{llr}
\tablecolumns{3}
\tablewidth{0pc} 
\tablecaption{Astrometric Noise Sources \label{noiseTermSummaryTable}}
\tablehead{ 
\colhead{Source} & \colhead{Section} & \colhead{Typical Magnitude ($\muas$)}}
\startdata
Temporal decoherence    & \ref{sec:temporal} & $\sim 5$  \\
Anisoplanatism          & \ref{sec:aniso}    & $0.2$     \\
Photon noise            & \ref{sec:photon}   & $3$       \\
Differential dispersion & \ref{sec:diffd}    & $\sim 30$ \\
Baseline errors         & \ref{sec:bline}    & $ < 10$   \\
Fringe template         & \ref{sec:template} & $1$       \\
Scan rate               & \ref{sec:scanrate} & $1$       \\
Beam walk               & \ref{sec:beamwalk} & $1$       \\
Global astrometry       & \ref{sec:global}   & $\ll 1$   \\
Star spots              & \ref{sec:starspot} & $< 8$     \\
Stellar granulation     & \ref{sec:granule}  & $< 3$     
\enddata
\tablecomments{
Sources of astrometric noise vary in magnitude from tens of micro-arcseconds 
to sub-microarcsecond levels.  Differential dispersion depends on color 
difference between binary components; for many targets, this is nearly zero, 
but for extreme color differences, this can be hundreds of microarcseconds.  
This is not the case for any of the PHASES targets.  Photometric variability 
accompanies star spots, of a magnitude that is easily detected for astrometric 
signatures of $8\,\microas$ or larger.
}
\end{deluxetable}

\subsection{Astrometric Observational Noise}\label{obsNoise}

In calculating the expected astrometric performance three major sources of 
error were taken into account:  errors caused by fringe motion during the 
delay sweep between fringes (loss of coherence with time), errors caused by 
differential atmospheric turbulence (loss of coherence with sky angle, 
i.e., anisoplanatism), and measurement noise in the fringe position.  Each is 
quantified in turn below, and the expected measurement precision is the 
root-sum-squared of the terms.

\subsubsection{Loss of Temporal Coherence}\label{sec:temporal}

The power spectral density of the fringe phase of a source observed through 
the atmosphere has a power-law dependence on frequency; at high frequencies 
typically
\begin{equation}
A(f) \propto f^{-\alpha}
\end{equation}
\noindent where $\alpha$ is usually in the range 2.5--2.7.  The effect of 
phase-referencing is to high-pass filter this atmospheric phase noise.  For 
PHASES, the servo was an integrating servo with finite processing delays and 
integration times, with the residual phase error ``fed forward'' to the second 
beam combiner \citep{lc03}.  The response of this system to an input 
atmospheric noise can be written in terms of frequency as 
\begin{eqnarray}
H(f)  & = &  \frac{1 - 2 {\rm sinc}(\pi f T_s)\cos(2 \pi f T_{d}) + {\rm sinc}^2(\pi f T_s) }{1 - 2 \frac{f_c}{f} {\rm sinc}(\pi f T_{s}) \sin(2 \pi f T_{d}) + \left ( \frac{f_c}{f} \right )^2 {\rm sinc}^2(\pi f T_{s})}
\end{eqnarray} 
\noindent where ${\rm sinc}(x) = \sin(x)/x$, $f_c$ is the closed-loop bandwidth
of the fringe-tracker servo (for PHASES $f_c = 10$ Hz or $5$ Hz), $T_s$ is the 
integration time of the phase sample (in standard mode, 6.75 ms), and $T_d$ is 
the delay between measurement and correction (done in post-processing, 
effectively 5 ms).  The phase noise superimposed on the double fringe measured 
by the astrometric beam combiner has a spectrum given by $A(f)H(f)$.

The sampling of the double fringe packet took a finite amount of time, first 
sampling one fringe, then the other.  In the time domain the sampling function 
can be represented as a ``top-hat'' function convolved with a pair of delta 
functions (one positive, one negative).  The width of the top-hat is equal to 
the time taken to sweep through a single fringe, while the separation between 
the delta functions is equal to the time to sweep between fringes.  In the 
frequency domain, this sampling function becomes
\begin{equation}
S(f) = \sin^2(2 \pi f \tau_p){\rm sinc}^2(\pi f \tau_\ast), 
\end{equation}
\noindent where $\tau_p$ is the time taken to move the delay between stars, 
$\Delta d/v_s$, and $\tau_{\ast}$ is the time to sweep through a single 
stellar fringe, $\Lambda/v_s$. $v_s$ is the delay sweep rate.

The resulting error in the astrometric measurement, given in radians by 
$\sigma_{tc}$, can be found from
\begin{equation}
  \sigma_{tc}^2 = \left (\frac{\lambda}{2\pi {\rm B}} \right)^2 \frac{1}{N}\int_0^\infty A(f)H(f)S(f) df
\end{equation}
\noindent where ${N}$ is the number of measurements.  It is worth noting that 
if phase-referencing was not used to stabilize the fringe, i.e.~$H(f) = 1$, the 
atmospheric noise contribution increases by a factor of $\approx 10^2$--$10^3$.
For $\sim$Hz scanning, the corresponding error is less than 10 $\muas$.

\subsubsection{Anisoplanatism}\label{sec:aniso}

The performance of a simultaneous narrow-angle astrometric measurement has 
been thoroughly analyzed by \cite{shao92}.  Here the primary result for the 
case of typical seeing at a site such as Palomar Mountain is restated, with 
the astrometric error in arcseconds due to anisoplanatism ($\sigma_{a}$) is
given by
\begin{equation}
\sigma_{a} = 540 B^{-2/3} \theta t^{-1/2} 
\end{equation}
\noindent where $B$ is the baseline (in meters), $\theta$ is the angular 
separation of the stars (in radians), and $t$ is the integration time in 
seconds.  This assumes a standard \citep{l80} atmospheric turbulence profile; 
it is likely that particularly good sites will have somewhat (factor of two) 
better performance.  For $\theta = 5\times 10^{-7} \sim 100$ mas, and 1 hr of 
integration, this term is roughly a fifth of a microarcsecond.

\subsubsection{Photon Noise}\label{sec:photon}

The astrometric error due to photon-noise ($\sigma_{p}$) is given in radians 
as  
\begin{equation}
\sigma_{p} = \frac{\lambda}{2\pi {\rm B}} \frac{1}{\sqrt{N}} \frac{1}{{\rm S/N}},
\end{equation}
\noindent where $N$ is the number of fringe scans, and S/N is the 
signal-to-noise ratio of an individual fringe.  For typical PHASES targets and 
PTI's throughput ($N\sim 2000$, $S/N \sim 10$), this was of order a few 
microarcseconds.

\subsection{Astrometric Instrumental Noise}

There were several effects internal to the instrument that could 
contribute noise 
terms or biases to the astrometric measurements.  Some could potentially vary 
on night-to-night timescales as the optical alignments vary on roughly these 
timescales.  Others resulted from properties of the measurement design.

\subsubsection{Differential Dispersion}\label{sec:diffd}

\begin{figure}[!ht]
\plotone{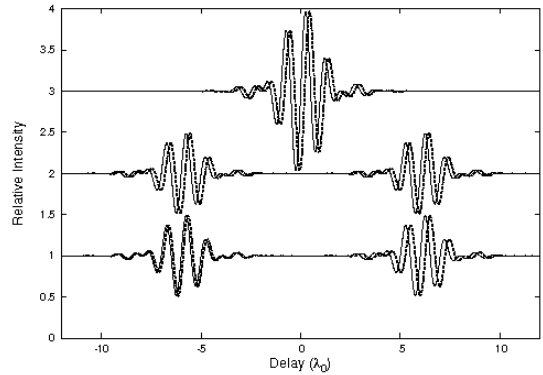}
\caption[Differential Dispersion] 
{ \label{fig:dispersion}
Schematic of the shift in fringe positions due to dispersion (the effect has 
been exaggerated for clarity).  The vacuum (no dispersion) interferograms are 
plotted with solid lines; those dispersed by air with dotted lines.  Top: 
dispersion shifts the point of zero optical path difference for a star, due to 
different amounts of air path in each arm of the interferometer (the effective 
optical path difference measured as if in vacuum).  Middle: the dispersion 
shifts for stars of equal colors were 
equal and canceled; the measured separation 
is the same.  Bottom: stars of unequal colors are shifted by slightly 
different amounts by dispersion, and the resulting measured separation is 
different.  For very extreme color differences, the shift can be hundreds of 
microarcseconds.  Not shown are the shape distortions to interferograms.
}
\end{figure}

The path compensation for the geometric delay at PTI was done with DLs 
in air.  At near-infrared wavelengths, air introduces a wavelength-dependent 
index of refraction given by 
\cite{allenquantities}
\begin{eqnarray}
n & = & 1 + \left(\frac{p T_s}{p_s T}\right)\left(
6.4328\times10^{-5} + 
\frac{0.029498}{146 - 1/\lambda^2} + 
\frac{2.554\times10^{-4}}{41 - 1/\lambda^2} 
\right)\nonumber\\
& & - 4.349\times10^{-5}\left(
1-7.956\times10^{-3}/\lambda^2
\right)\frac{p_w}{p_s}
\end{eqnarray}
\noindent where $\lambda$ is the vacuum wavelength in $\rm{\mu m}$, $p$ is the 
air pressure, $p_w$ is the partial pressure of water vapor, 
$p_s = 1.01325\times10^{5}$ Pa, $T$ is the temperature, and $T_s = 288.15$ K.  
The fringe packets of astrophysical sources were dispersed by an amount that 
depends on the difference in air paths between arms of the interferometer; 
this changed the shape and overall location of the fringe packets; see 
Figure \ref{fig:dispersion}.  If two 
stars were in the same beam and identical in color, the change in location 
was common to both and canceled; similarly, the distortions of the fringe 
packets are common and canceled to first order in a differential measurement.  
Simulated data and reanalysis of real data with modified fringe templates 
showed that higher order terms in the fringe packet shape are negligible at 
the microarcsecond level.

If, however, the two stars were of differing colors, each would be dispersed by 
a slightly different amount, and their apparent separation would be biased.  
The shift in the apparent position of each star's fringes can be approximated 
by evaluating the dispersion at the effective average wavelength of the star in 
the passband.  The effective average wavelength was calculated by multiplying 
the instrumental bandpass by the stellar spectrum.  For an order-of-magnitude 
estimate of the effect of differential dispersion, one can model the 
instrumental bandpass as a tophat function passing wavelengths 
$2-2.4 {\rm \mu m}$ (nominal K-band) and the stellar spectra as blackbodies.  
The shifts in apparent positions for several spectral types over 40 m of 
differential air path (a maximum amount for PTI) are given in Table 
\ref{diffDispTable}.  Note that for G5-K5 binaries, the amount is 35.8 
$\microas$ and for B5-A5 it is 30.6 $\microas$.  For much more extreme color 
ratios, the effect can be as large as 150.6 $\microas$ for B5-M5 binaries; the 
PHASES sample did not include such systems, as their high contrast ratios 
prevented observation.

\begin{deluxetable*}{lrrrrrr}
\tablecolumns{7}
\tablewidth{0pc} 
\tablecaption{Differential Dispersion \label{diffDispTable}}
\tablehead{ 
\colhead{Spectral} & \colhead{Effective} & \colhead{Effective K-band} &
\colhead{$\left(n-1\right)$} & \colhead{$\left(n-n_{\rm F5}\right)$} & 
\colhead{$\left(n-n_{\rm F5}\right)$} & \colhead{Error vs.~}\\
\colhead{Type} & \colhead{Temperature (K)} & 
\colhead{Wavelength $({\rm \mu m})$} & \colhead{$\times 10^4$} & 
\colhead{$\times 10^9$} & 
\colhead{$\times38 {\rm m}$ $\left[ {\rm nm} \right]$ } &
\colhead{F5 $\left[ \microas \right]$}}
\startdata
O5 & 44500 & 2.1763 & 2.729232 & 0.93  & 35.4 & 73   \\
B5 & 15400 & 2.1772 & 2.729229 & 0.66  & 25.0 & 51   \\
A5 & 8200  & 2.1785 & 2.729225 & 0.25  & 9.7  & 20   \\
F5 & 6440  & 2.1794 & 2.729223 & 0.00  & 0.0  & 0    \\
G5 & 5770  & 2.1799 & 2.729221 & -0.14 & -5.4 & 11   \\
K5 & 4350  & 2.1815 & 2.729217 & -0.61 & -23.3 & 48  \\
M5 & 3240  & 2.1839 & 2.729210 & -1.32 & -50.3 & 103 
\enddata
\tablecomments{
Effect of color-dependent differential dispersion.  Stellar temperatures are 
for dwarf stars, from \cite{CarrollOstlie}.  All numbers are for zero water 
vapor pressure, $p_w = 0$.  Increasing water vapor pressure to $p_w = p_s$ 
increases the astrometric effect by a factor of roughly $20\%$.
}
\end{deluxetable*}

Because the stars were often observed at the same hour angles from one night to 
the next (and thus the delay positions are relatively common between nights), 
this effect introduced a much smaller scatter than that listed in the table.  
However, it may have introduced 
biases in the stellar separations and introduced 
scatter between observations taken in multiple baselines (for which the delay 
positions differed).  These biases and scatters were of order the amounts given 
in Table \ref{diffDispTable}.

The binaries in the PHASES sample were generally of components with equal 
brightnesses and thus similar colors.  No hour-angle dependent biases 
significant on the level of the precision of the observations were observed.  
This effect is likely to be important for traditional narrow angle astrometry 
methods at the Keck Interferometer or Very Large Telescope Interferometer, 
which aim to use field stars as astrometric references for nearby stars, and 
reference and target will often have very different colors.  The effect is 
largely reduced if one uses a spectrometer to measure the group delay positions 
of the fringes.

Starting in 2005, the longitudinal dispersion compensator addressed this 
potential systematic.  The measurements acquired with this correction are 
flagged in Table \ref{phasesAstrometryData}.

\subsubsection{Baseline Errors}\label{sec:bline}

The baseline vector used in the differential delay equation to determine 
astrometric quantities was derived by inversion of the delay equation from the 
fringe locations of point-like sources with known global astrometric 
positions.  Uncertainties and variability of the baseline vector were sources 
of differential astrometry uncertainties via the differential delay equation.  
An incorrect baseline model would show up as an hour-angle dependent error 
term that would potentially increase night-to-night scatter beyond that 
predicted by the formal uncertainties; this was tested by dividing data sets 
within single nights into multiple sets by hour-angle range and the results 
were self-consistent.

\begin{figure*}[!ht]
\plotone{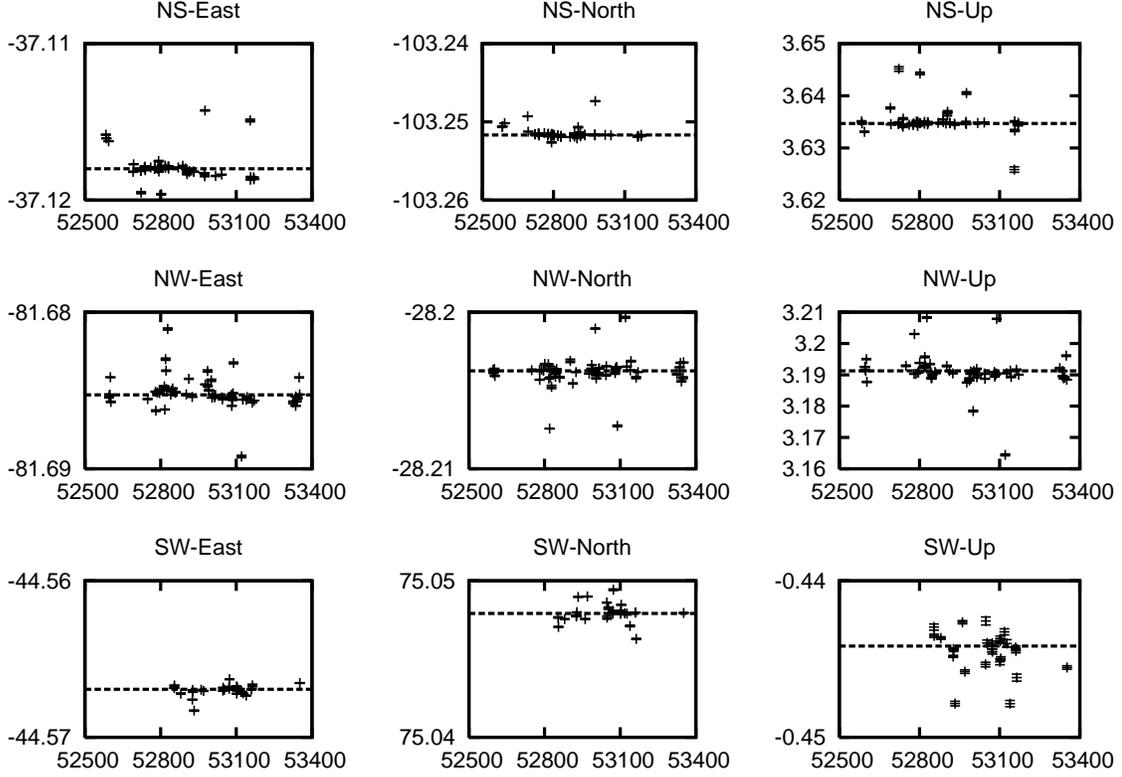}
\caption[Baseline Repeatability] 
{ \label{fig:baselineRepeatability}
Solutions for the three PTI baseline vectors.  The three baselines at PTI were 
named ``NS'', ``NW'', and ``SW'' due to their rough orientations.  Each was a 
three dimensional vector, which was given by components in the ``East'' 
(east-west), ``North'' (north-south), and ``Up'' (up-down) directions (the 
first two were tangent to the Earth, the last was perpendicular).  Horizontal 
axes are time in modified Julian Days (MJD), vertical axes are baseline length 
in meters.  Lines represent average baseline fits used for data reduction 
presented in this paper; points with error bars represent a given night's 
baseline solution.  The baseline solutions were derived from the observed delay 
positions of single-star sources with known global astrometric positions via 
inversion of Equation \ref{delayEquation}.  The y-axis tick marks in each plot 
are all 10 mm.  Note that the scatter in the ``Up'' dimension was much larger 
than the other dimensions; this was due to preferential observing of targets 
overhead, for which the ``Up'' component is highly covariant with the constant 
term in the delay equation.  The baseline solution used for data analysis was a weighted average 
of the solutions plotted.
}
\end{figure*}

No evidence of hour-angle dependent error terms was seen in the PHASES data, 
supporting evidence that the baseline models are correct.  As shown in Figure 
\ref{fig:baselineRepeatability}, except for a few outliers (likely due to 
using point sources with poor global astrometry values or a night's 
observation only covering a small range of hour angles or declinations) the 
night-to-night drift in baseline model solutions were less than 1 mm in 
north-south and east-west directions for the two primary baselines used for 
PHASES observations (NS and SW).  The up-down dimension was stable to a few 
millimeters in both cases; this scatter was likely due to limited measurement 
precision rather than actual baseline variability, implying that it could be 
improved by averaging several nights' values.

The amount by which a baseline error of $\overrightarrow{\sigma_{B}}$ affects 
a differential astrometry measurement $\overrightarrow{\Delta S}$ was 
determined as follows.  To maintain the same observed differential delay 
between stars, the differential delay equation requires that
\begin{equation}
\overrightarrow{B}\cdot\overrightarrow{\Delta S} = \left(\overrightarrow{B} + \overrightarrow{\sigma_{B}}\right)\cdot\left(\overrightarrow{\Delta S} + \overrightarrow{\sigma_{\Delta S}}\right)
\end{equation}
\noindent where $\overrightarrow{\sigma_{\Delta S}}$ is the astrometric error 
caused by baseline error $\overrightarrow{\sigma_{B}}$.  Canceling like terms 
and assuming 
$\overrightarrow{\sigma_{\Delta S}}\cdot\overrightarrow{\sigma_{B}}$ is less 
than the other terms simplifies this to
\begin{equation}\label{baselineVector2}
\overrightarrow{\sigma_{B}}\cdot\overrightarrow{\Delta S} = - \overrightarrow{B}\cdot\overrightarrow{\sigma_{\Delta S}}.
\end{equation}

The vector $\overrightarrow{\Delta S}$ was tangent to the celestial sphere; 
only that component which was not perpendicular to the baseline was actually 
measured (this measured component of the separation is referred to as 
$\delta S$) and only its uncertainty ($\sigma_{\delta S}$) was thus applicable.  
The angle between these measured components and the baseline vector is given 
by the target's zenith angle $z$; this was always kept to less than 
$45^\circ$.  Of course, the baseline uncertainty vector 
$\overrightarrow{\sigma_{B}}$ need not be oriented with $\overrightarrow{B}$ 
itself; its components $\sigma_{Bx}$ and $\sigma_{By}$ tangent to and 
$\sigma_{Bz}$ normal to the Earth (also referred to as the ``U'' or ``Up'' 
component) are introduced.  Substituting into Equation \ref{baselineVector2} 
gives the relationship between baseline error and astrometric error as
\begin{equation}
\delta S \left(\left(\sigma_{Bx}\cos\phi + \sigma_{By}\sin\phi\right)\cos z + \sigma_{Bz}\sin z\right) = 
- \left|B\right|\sigma_{\delta S}\cos z, 
\end{equation}
\noindent where $\phi$ is an angle determined by the hour angle and 
declination of the target.  On rearranging terms, the fractional astrometric 
measurement uncertainty due to baseline uncertainties was
\begin{equation}
\frac{\sigma_{\delta S}}{\delta S} = - \frac{\sigma_{Bx}\cos\phi + \sigma_{By}\sin\phi + \sigma_{Bz}\tan z}{\left|B\right|}.
\end{equation}
\noindent  For $\left| B \right| = 100$ m and $z < 45^\circ$, baseline 
uncertainties of 2 mm cause 10 $\microas$ errors in the astrometry for a 
binary with projected separation $\delta S = 0.5$ arcsec.  Though the 
measured component of $\Delta S$ continually varies as the Earth rotates the 
baseline vector, the above derivation is true at any given instant.  Earth 
rotation causes errors to appear in both astrometric dimensions.

A slightly more subtle baseline effect was that there can be differences 
between the wide-angle (``astrometric'') and narrow-angle (``imaging'') 
baselines.  When determining the baseline vector via the delay positions for 
stars with known global astrometry solutions spread across the sky, the 
siderostats are repointed for each observation to place the target star 
on-axis.  Thus, one was measuring the vector between the pivot points of the 
siderostats, about which the siderostats were repointed 
for each target.  When measuring the 
separations between two stars in a single field with a single pointing of the 
telescopes, at least one of the stars would be off axis; one instead desired 
to know the separations between the corresponding points on the surfaces of 
the two siderostats to get the proper 
delay scale.  Through careful optical design, PTI was built to minimize 
differences between the astrometric and imaging baselines in order that the 
original narrow angle astrometry mode would function at the 100 $\muas$ level 
for binaries with separations of tens of arcseconds.  Because errors due to 
baseline uncertainty 
scale with binary separation, the effect was negligible at the level of a few 
microarcseconds for subarcsecond binaries in the PHASES sample.

\subsubsection{Fringe Template}\label{sec:template}

Because the astrometric measurements were differential between two stars, 
they were 
relatively insensitive to the model fringe template.  The fringe model 
used in the astrometric analysis was determined by observing interferograms of 
single stars.  An effective bandpass was constructed from an incoherent 
averaging of the periodograms of many such interferograms, and used to 
recompute a standard interferogram template, to be applied to the data.  This 
effective bandpass was only an approximation for most stars, as there were 
variations in source temperature and spectra.  However, reanalysis with 
several different fringe models showed variations only at the single 
microarcsecond level.

\subsubsection{Scan Rate and Earth Rotation}\label{sec:scanrate}

Earth rotation caused variable projection of the binary separation on the 
interferometer baseline vector.  The details of the variability depend of the 
observatory location, sky position of the target binary, and the orientation 
of the baseline vector, but for order-of-magnitude estimations, can be 
approximated as a sinusoid with period of 1 day and amplitude equal to the 
total binary separation $a$:
\begin{equation}
\Delta s \approx a \times \cos \left( 2 \pi t / {\rm day} \right).
\end{equation}
\noindent The differential delay rate was given by the first derivative of this 
equation with respect to time, converted from sky angle to delay length by the 
interferometer's resolution.  For $a \sim 500$ mas, this differential delay 
rate is about 20 ${\rm nm \, s^{-1}}$, or 5 nm (10 $\microas$) in the 
(typically) 250 ms required to scan between the fringe packets.  Roughly an 
equal 
number of scans were obtained in each scan direction (to within $10\%$), and 
this effect canceled to first order (to the same level, $10\%$ or 1 
$\microas$).  However, curvature in the differential delay motion does not 
cancel; it is given by the second time derivative of the projected separation 
and was roughly $1.4\times10^{-3}\,{\rm nm\,s^{-2}}$ (less than 3 
nano-arcseconds $s^-2$).  Thus the differential delay rate was small 
enough, and the measurement rate fast enough, that the finite measurement rate 
did not contribute significant uncertainties.

\subsubsection{Beam Walk}\label{sec:beamwalk}

\begin{figure}[!htp]
\plotone{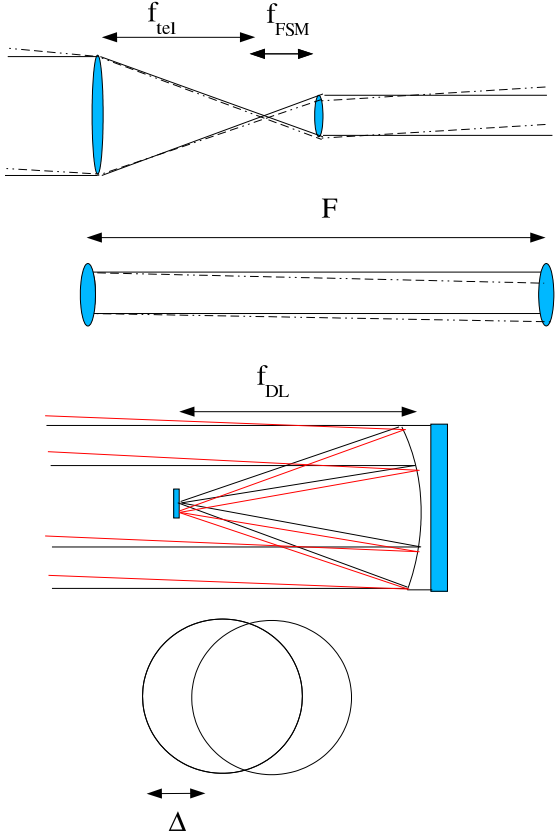}
\caption[Beam Walk]
{ \label{fig:beamwalk}
Three instances where beam walk could occur, causing stars at slightly different 
sky angles to illuminate different parts of optical elements.  Top: shear 
introduced at the telescope by focusing and recollimating the beam.  ``FSM'' 
stands for the ``Fast Steering Mirror'', which provided tip-tilt corrections 
(first-order adaptive optics) and recollimated the light after the telescope.  
Second from top: shear within a collimated beam over large optical paths.  
Second from bottom: shear at focus of DL optics (the movable 
mirrors that provide optical delays).  Bottom: the shear of two beams by 
amount $\Delta$, causing only partial overlap.
}
\end{figure}

The interferometer telescopes imaged a sky field and then recollimated the 
light.  Through this process, light from two stars separated on the sky by 
angle $\alpha$ was partially sheared with respect to each other and 
proceeded to illuminate slightly different parts of the optics that guide the 
light to the detector.  Starlight in a recollimated beam that originated from 
different sky positions also developed relative shear equal to the path 
traveled multiplied by their angular separation (see Figure 
\ref{fig:beamwalk}).  To the extent that the optics were imperfect (had rough 
surfaces), the light from each star traveled slightly different path-lengths 
from telescope to detector.  This process is known as beam walk.

\cite{ColavitaBW} determined the extent to which beam walk 
introduces astrometric errors.  That approximate stochastic analysis yielded 
an rms pathlength error $\epsilon$ given by
$$
\epsilon \sim w \left ( \frac{\delta}{q} \right ) \left ( \frac{q}{z}
\right )^{1/4}, \delta \ll q,
$$
where $w$ is the rms wavefront error over an optic of diameter $z$, $q$ is the 
diameter of the starlight footprint, and $\delta$ is the linear beamwalk.  For 
optics with a surface quality of $\lambda/20$ peak-to-valley (like 
those used at PTI), the wavefront rms is $w \simeq \lambda/40$.  The actual 
error is $\sim \sqrt{2}$ larger than this due to the presence of two 
telescopes in the interferometer.

The first place where beam walk could occur 
was within the telescope itself.  The 
beam was collimated at the telescope primary, focused by the primary, and 
recollimated to a $q = 0.075$ m beam by a $z = 0.1$ m diameter mirror to be 
fed to the ``Fast Steering Mirror'' (FSM; this mirror corrected for tip-tilt 
wavefront errors 
across the telescope, providing low-order adaptive optics).  The distance from 
primary mirror to this mirror was the sum of their focal lengths, 4.75 m.  The 
beam walk over 0.1 arcsec ($4.8 \times 10^{-7}$ rad) was thus 
$\delta = 2.3\,\mu{\rm m}$.  Beam walk on the FSM 
thus contributed an astrometric error of 0.001 $\microas$ and was negligible.

The relative angles of starlight in the recollimated beam were increased by a 
factor of the ratio of the primary mirror and FSM focal lengths (5.33), thus 
light from sky locations separated by 0.1 arcsec had a differential angle 
of 0.533 arcsec ($2.6 \times 10^{-6}$ rad).  This recollimated beam 
from the FSM traveled through light pipes to the beam combining laboratory, 
where movable mirrors added 
a variable amount of delay.  This total travel is of 
order 50 m; the mirrors are typically $z = 0.1$ m diameter.  The 
beam walk over 0.533 arcsec was $\delta = 130\,\mu{\rm m}$, 
which contributed an astrometric error of 
0.07 $\microas$.  There were a few mirrors along this path, and the total 
astrometric error would be determined by considering the optical qualities of 
all optics and adding the effects in quadrature.  Because this beam walk was 
so small, the sum total of these remained negligible.

The movable mirrors were comprised of a parabolic mirror of focal length 
1.07 m and a small ($z \approx 0.01$ m) flat mirror located near its focus.  
Collimated 
light was directed to one side of the parabola, focused onto the flat mirror, 
then recollimated by the parabola's other side.  On the flat mirror, the 
(diffraction-limited) beam diameter was only 
$q \sim \lambda f / d = 31\,\mu{\rm m}$ where $\lambda$ is the operating 
wavelength of light, $f$ is the parabola's focal length, and $d$ is 
the collimated beam diameter (0.075 m).  The beam walk was 
$2.8\,\mu{\rm m}$.  The flat mirrors 
contributed an astrometric error of 0.9 $\microas$ from beam walk.  It is 
concluded that beam walk did not contribute significant measurement errors.

\vspace{0.5in}

\subsubsection{Global Astrometry Errors}\label{sec:global}

Uncertainty in the global position of a target binary on the celestial sphere 
was coupled into the differential astrometric measurement.  Errors in right 
ascension were equivalent to measurement timing errors; declination 
uncertainties had similar effects.  The order of magnitude of this effect can 
be derived as follows:  the fractional error in global astrometry (error in 
arcseconds divided by total number of arcseconds around a quarter circle) is 
roughly equal 
to the fractional error in differential astrometry separation vector 
(astrometric error divided by binary separation).  A 1 arcsec global 
astrometry error caused differential astrometric errors of less than 3 
$\microas$ for binaries of separation 1 arcsec or less.  Typical 
uncertainties in global astrometry were much less than an arcsecond, with 10 
mas being a much more common value.  Effects such as stellar 
aberration (20 arcsec) were accounted for in the PHASES data reduction 
software; if ignored, these could cause significant differential astrometry 
uncertainties.

\subsection{Astrometric Astrophysical Noise}

There were potential sources of apparent astrometric motion in the target stars 
due to processes within the stars 
themselves.  These included star spots and stellar granulation.

\subsubsection{Star Spots}\label{sec:starspot}

As previously presented by \cite{Mut06_v819her}, the maximum shift in the 
center-of-light of a star caused by star spots can be evaluated with a model 
comprised of a uniform stellar disk (radius $R$) except for a zero-temperature 
(non-emitting) circular region of radius $r$ tangent to the edge of the 
stellar disk (i.e.~centered at $x=R-r$, $y=0$).  The center of light is 
displaced by:
\begin{eqnarray}\label{spot_ast}
\frac{x_c}{R} & = & \frac{\int_{-R}^{R} \int_{-\sqrt{R^2-x^2}}^{\sqrt{R^2-x^2}} x {\rm d} y {\rm d} x - 
\int_{R-2r}^{R} \int_{-\sqrt{r^2-\left(x-R+r\right)^2}}^{\sqrt{r^2-\left(x-R+r\right)^2}} x {\rm d} y {\rm d} x}
{R\pi\left(R^2-r^2\right)}\nonumber\\
& = & -\frac{r^2/R^2}{1+r/R}.
\end{eqnarray}

The presence of star spots can be confirmed through photometric measurements 
simultaneous with astrometric observations.  The non-emitting spots in this 
model would cause photometric variations proportional to the fractional area 
of the stellar disk covered:
\begin{equation}\label{spot_phot}
\frac{F}{F_o} = 1-\frac{r^2}{R^2}
\end{equation}
\noindent where $F_o$ is the star's flux when no spots are present.  Equations 
\ref{spot_ast} and \ref{spot_phot} provide a relationship between the apparent 
astrometric and photometric shifts caused by star spots.  

The largest possible astrometric shift by a star spot is given by evaluating a 
slightly different model.  In this case, the star spot fills the 
(non-circular) area from the star's edge to a chord at distance $x_o$ from the 
star's true center.  The astrometric shift is 
\begin{eqnarray}\label{spot_ast2}
\frac{x_c}{R} & = & \frac{\int_{-R}^{x_o} \int_{-\sqrt{R^2-x^2}}^{\sqrt{R^2-x^2}} x {\rm d} y {\rm d} x}
{R \int_{-R}^{x_o} \int_{-\sqrt{R^2-x^2}}^{\sqrt{R^2-x^2}} {\rm d} y {\rm d} x}\nonumber\\
& = & -\frac{2\left(1 - \frac{x_o^2}{R^2}\right)^{3/2}}
{3 \left( \frac{\pi}{2} + \arcsin\frac{x_o}{R} + \frac{x_o}{R}\left(1 - \frac{x_o^2}{R^2}\right)^{1/2}\right)}
\end{eqnarray}
\noindent with corresponding photometric variations of 
\begin{equation}\label{spot_phot2}
\frac{F}{F_o} = \frac{1}{\pi}\left(\frac{\pi}{2} + \arcsin\frac{x_o}{R} + \frac{x_o}{R}\left(1 - \frac{x_o^2}{R^2}\right)^{1/2}\right).
\end{equation}

\begin{figure}[!ht]
\plotone{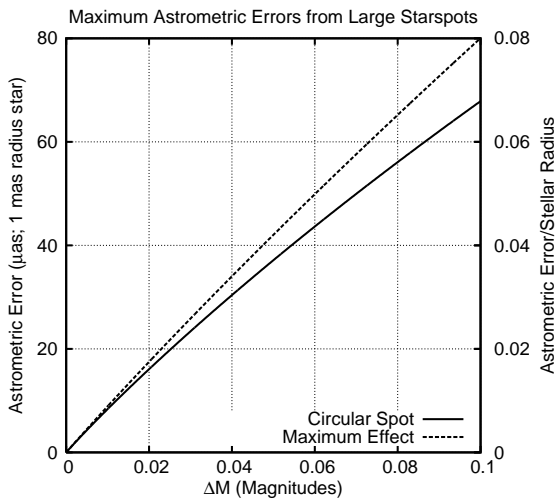}
\caption[Star Spots and Astrometry] 
{ \label{fig:starspot_ast}
Maximum effect of star spots on astrometric measurements vs.~the 
photometric variations they cause.}
\end{figure}

For stars of typical radius 1 mas, the simplified model gives a 
roughly linear relationship of 0.8 $\microas$ of astrometric shift per 
milli-magnitude of photometric variability, see 
Figure \ref{fig:starspot_ast}.  Only stars with diameters smaller than PTI's 
resolution of $\sim 4$ mas were observed by PHASES.  
Photometric variations of these 
scales can be monitored by small ground-based telescopes.  The timescale of 
these variations is on order the rotation rate of a star (days to weeks).  
\cite{lanza2008} re-evaluated the effects of spots with a model that includes 
stellar limb darkening and lower spot contrast (``warm'' spots rather than the 
completely non-emitting spots assumed above) and found this reduces the 
astrometric effect by $\sim 40\%$.

\subsubsection{Stellar Granulation}\label{sec:granule}

Stellar granulation causes photometric variability of subsections of a star's 
surface.  Averaged over the whole of the stars surface, these photometric 
variations can cancel to large extent and the intrinsic variability of the 
star remains small, though with a large astrometric uncertainty.  
\cite{Svensson2005} showed that the effects of stellar granulation are 
independent of a star's radius but are strongly correlated with surface 
gravity, and provide values for the astrometric effects in white light.

For stars with very low surface gravities (i.e.~red giants), astrometric 
perturbations can be quite large---as much as $300\,\microas/D\,\rm{[pc]}$.  
Red giants within 100 pc were overresolved by PTI and could not be 
observed, thus even for these stars this effect was negligible.  For 
main-sequence stars, the effect was closer to $0.1\,\microas/D\,\rm{[pc]}$.

\section{Empirical Error Bar Scaling Rules}\label{errorBarCorrection}

The formal uncertainty ellipses derived with the procedure outlined in 
Section \ref{sec::reduction} were found to underestimate the actual scatter 
of the measurements to an orbital model.  Two reasonable approaches 
to correct this are to find a multiplicative scale factor by which to 
increase the formal uncertainties, or a noise floor to be 
added-in-quadrature to the formal uncertainties.  Unfortunately, no 
single multiplicative factor nor noise floor was found that could bring 
the measurement uncertainties and fit residuals into satisfactory 
agreement for a majority of the stars.  The instrumental changes 
made in 2005 to introduce a longitudinal dispersion compensator (discussed 
in Section \ref{sec:diffd}) and an automated alignment system further 
complicate this effort.  However, upon dividing the measurements into two 
subsets according to whether or not the instrumental changes were implemented 
(the subset without the upgrades will be 
referred to as subset 1, that with the 
upgrades will be referred to as subset 2), 
and allowing for different scaling or noise floor factors in each subset, 
again no single solution was obtained.

Recognizing that both multiplicative factors and noise floors might be 
present, and could 
affect the major and minor axes of the uncertainty ellipses in different 
ways, a solution was sought to incorporate parameters for both types of terms, 
allowing for different values in the two axes and in the two subsets.  This 
finally produced a single satisfactory solution.  The procedure used to find 
the solution and the final values obtained were as follows.

Of the 51 star systems observed by PHASES, 33 had 10 or more measurements.  
Six of those 33 had been previously discovered to be triple or 
quadruple star systems, and a seventh (HR 2896 $=$ HD 60318) 
was discovered by PHASES 
to be a triple system; for simplicity of analysis, the effort to derive 
the uncertainty scaling rules was limited to binaries.  Three more systems 
had too few measurements in one or both subsets 1 and 2, and were not included 
in the analysis.  Of the remaining 23 binaries, 15 ($\sim 2/3$) were selected 
for evaluation; the other $\sim 1/3$ of the systems were not included allowing 
for the possibility that their increased levels of scatter were the result of 
astrophysical phenomena; however, the empirical scaling rules have been found 
to work satisfactorily on many of these as well, as a check of the rules' 
validities.

The size of the major and minor axes of each measurement was resized by the 
formula
$$
\sigma_{(i, j), c}^2 = f_{(i,j)}^2\sigma_{(i,j), {\rm raw}}^2 + \sigma_{(i,j), q}^2
$$
where $i=1, 2$ indicates the subset number, $j=1, 2$ indicates whether it is the 
uncertainty ellipse major (1) or minor (2) axis, 
$\sigma_{(i, j), c}$ is the ``corrected'' 1-$\sigma$ uncertainty, $f_{(i,j)}$ is 
a multiplicative scaling factor, $\sigma_{(i,j), {\rm raw}}$ is the formal 
1-$\sigma$ uncertainty, and $\sigma_{(i,j), q}$ is a noise floor added in 
quadrature.

An iterative procedure was used to optimize the values of the eight parameters 
$f_{(i,j)}$ and $\sigma_{(i,j), q}$ in order to produce the most statistically 
reasonable set of uncertainty ellipses to represent the observed scatter about 
Keplerian models for the 15 stars.  The constraints for optimizing those eight 
parameters were:  (1) within each subset, for each axis, the $\chi_r^2$ 
goodness-of-fit statistics computed along that axis should have minimal variance 
among the binaries (i.e.~all the binaries being evaluated should have similarly 
good fits in a given axis), (2) averaged over all the binaries, the $\chi_r^2$ 
statistic computed within each subset/axis combinations should be 
equal to the other subset/axis (and unity; i.e.~fit residuals should equally 
distributed among the various subsets and axes), and (3) the fits for as many 
binaries as possible should have $\chi_r^2$ as close to unity as possible.

The eight parameters $f_{(i,j)}$ and $\sigma_{(i,j), q}$ were given nominal 
values, and the 15 binaries' PHASES measurements were fit to Keplerian models 
using the replacement uncertainty ellipses.  The correction parameters were 
varied one at a time, each step recalculating the Keplerian models, until the 
variance of the $\chi_r^2$ among the binaries on the affected subset/axis 
was minimized (criterion 1).  This was repeated over the eight parameters, 
after which the average $\chi_r^2$ was calculated in each of the four subset/axis 
combinations, and the parameters affecting each subset/axis combination 
were renormalized to force the average $\chi_r^2 = 1$ for that combination
(criterion 2).  This procedure was then iterated, varying the parameters 
individually to minimize $\chi_r^2$ variance followed by renormalizing to 
$\chi_r^2 = 1$ for each axis, until the parameter values converged.

The final parameter values that produced the best universal uncertainty 
scaling rules were found to be:
\begin{itemize}
\item $f_{(1,1)} = 1.3$ (original instrument configuration, major axis); 
\item $f_{(1,2)} = 3.8$ (original instrument configuration, minor axis); 
\item $f_{(2,1)} = 1.3$ (revised instrument configuration, major axis); 
\item $f_{(2,2)} = 1.0$ (revised instrument configuration, minor axis); 
\item $\sigma_{(1,1), q} = 140 \, {\rm \mu as}$ (original instrument configuration, major axis); 
\item $\sigma_{(1,2), q} = 35 \, {\rm \mu as}$ (original instrument configuration, minor axis); 
\item $\sigma_{(2,1), q} = 140 \, {\rm \mu as}$ (revised instrument configuration, major axis);
\item $\sigma_{(2,2), q} = 35 \, {\rm \mu as}$ (revised instrument configuration, minor axis).
\end{itemize}
Only four non-trivial parameters end up being 
needed to correct the measurement 
uncertainties:  a single multiplicative factor of 1.3 for the major axis 
(independent of subset), a multiplicative factor of 3.8 for the minor axis 
of data that lacked the dispersion compensator and/or automatic alignment 
system ($f_{(2,2)} = 1$ is a trivial value representing no correction is 
needed), and noise floors of 140 and 
35 ${\rm \mu as}$ to be added in quadrature 
to the major and minor axes, respectively (independent of subset).  These 
values are not entirely surprising:  a large multiplicative factor for the 
minor axis before the instrument revisions disappeared as those changes made 
the instrument more stable and increased path monitoring.  A noise floor 
of 35 ${\rm \mu as}$ in the minor axis, regardless of subset, is reasonably 
expected from the many noise sources described 
in Section \ref{sec::expectedperf}.  
The root-sum-square of the errors in Table \ref{noiseTermSummaryTable} is 15 
$\microas$.  Given that some of the simple arguments presented were designed 
to provide error estimates at only the factor of 2 level, the resulting 35 
$\microas$ noise floor is not unreasonable.  In addition, it is possible some 
differential dispersion persisted even with the dispersion compensator in 
place, though it certainly reduced the amount.  
Because the major axis represents a direction 
perpendicular to that at which the 
interferometer baseline was oriented at the 
average measurement time, the measurement 
along that axis is slow to build via Earth-rotation 
synthesis, and it is not surprising 
to find a small multiplicative factor and 
larger noise floor in that direction.

After applying these corrections to the 
measurement uncertainties, the $\chi_r^2$ statistic 
was calculated for Keplerian fits to the 
data of all 26 binaries having more than 10 
observations.  Fourteen of the 26 (54\%) 
evaluate to $0.5 < \chi_r^2 < 1.5$, and an 
additional two just miss the cutoff (with values 
of 0.44 and 0.46; including these would bring 
the total to 62\%).  Furthermore, four of the systems with too-large values 
of $\chi_r^2$ are much better modeled by a double Keplerian model, indicating 
an additional component may be present; see Paper V.  The 26 binaries, number 
of observations, the fits' $\chi_r^2$ metrics, and 
which binaries were used to evaluate the scaling 
rules are listed in Table \ref{tab::scaling}.

\begin{deluxetable}{llll}
\tablecolumns{4}
\tablewidth{0pc} 
\tablecaption{Binaries Evaluated Using Empirical 
Uncertainty Ellipse Scaling Laws\label{tab::scaling}}
\tablehead{ 
\colhead{HD Number} & \colhead{Nights Observed} & 
\colhead{$\chi_r^2$} & \colhead{}}
\startdata
17904 & 46 & 0.53 & * \\
26690 & 20 & 0.75 &  \\
44926 & 23 & 0.62 & * \\
76943 & 16 & 1.10 &  \\
114378 & 24 & 1.24 & * \\
137107 & 51 & 1.22 & * \\
137391 & 23 & 0.59 & * \\
137909 & 73 & 0.85 & * \\
140159 & 15 & 1.15 & * \\
171779 & 54 & 1.04 & * \\
187362 & 10 & 0.97 &  \\
202275 & 69 & 0.91 & * \\
207652 & 50 & 0.58 & * \\
214850 & 53 & 0.87 & * \\
\hline
5286 & 18 & 4.71 &  \\
6811 & 20 & 2.16 &  \\
13872 & 89 & 0.46 & * \\
77327 & 47 & 0.44 & * \\
81858 & 12 & 2.27 &  \\
129246 & 17 & 1.86 & * \\
140436 & 43 & 1.82 & * \\
155103 & 10 & 0.21 &  \\
176051 & 66 & 4.43 &  \\
196524 & 73 & 2.68 &  \\
202444 & 39 & 3.52 &  \\
221673 & 99 & 8.01 &  
\enddata
\tablecomments{
Column 1 is the star's HD Catalog Number, Column 2 is the total number of nights the star 
was observed, and Column 3 is the value of $\chi_r^2$ calculated for a Keplerian model.  
An asterisk appears in the fourth column of stars used to evaluate the 
uncertainty ellipse scaling rules.
}
\end{deluxetable}

\vspace{0.2in}

\section{Distribution of Fit Residuals}\label{sec::residualPlots}

\begin{figure*}[!ht]
\plottwo{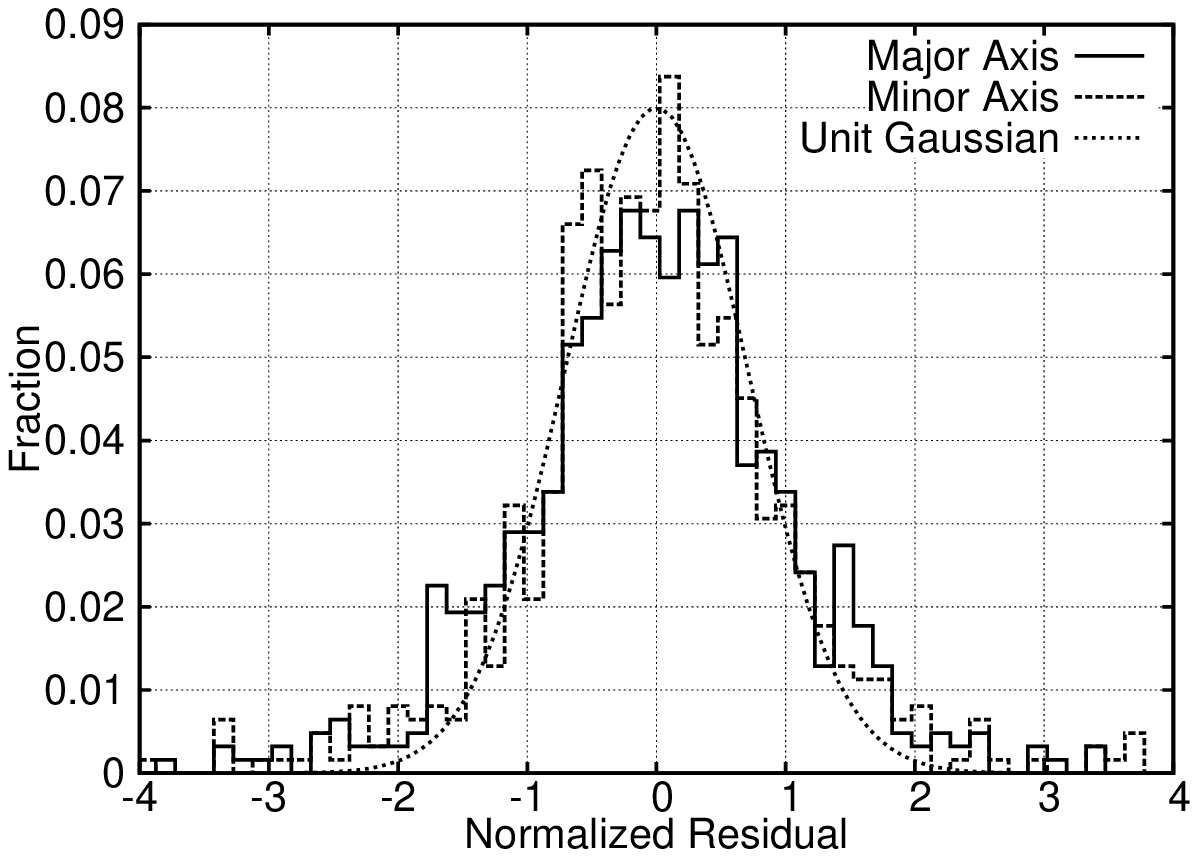}{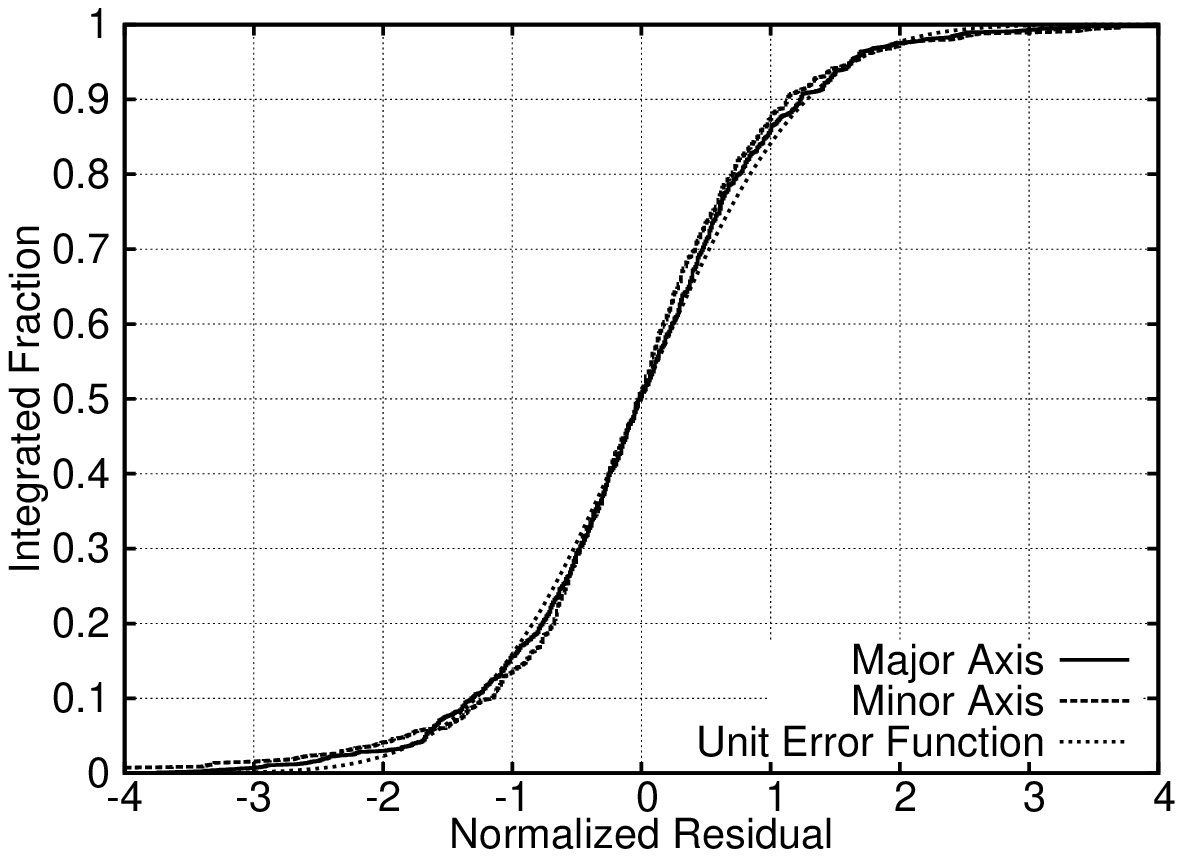}
\caption[Statistics of PHASES Data Residuals] 
{ \label{fig::residPlots}
Distributions of normalized residuals for two-body Keplerian fits to 
20 binaries, in major and minor axes.  On the left is the histogram of the 
residuals and unit-width Gaussian for comparison, while on the right 
is the continuous distribution function, with the theoretical distribution 
for Gaussian noise (the error function).
}
\end{figure*}

To use the PHASES measurements for orbit fitting and companion searches, it was 
important to establish the statistical properties of the measurement residuals 
to determine whether they are Gaussian.  Using uncertainties derived from 
the process outlined in Section \ref{errorBarCorrection}, 
Keplerian orbits were fit for 19 of the 20 
binaries presented in Paper II (HD 202444 has been omitted, since there is 
some evidence to suggest it hosts a substellar companion; see Paper V), as 
well as $\delta$ Equ (HD 202275), which has previously been studied at length 
\citep{Mut05_delequ, MuteMuOri2008}.  The fits included all the non-PHASES and 
radial velocity measurements as listed in Paper II, rather than just the PHASES 
measurements, to ensure the residuals represented the best known binary 
motions.  The PHASES measurement residuals along the 
minor axes of the error ellipses for these 20 binaries 
were normalized by their minor axis measurement uncertainties and 
combined into a joint residual distribution; a similar distribution was made 
for the major axis residuals.  Histograms and continuous distribution functions 
(the latter being the integral of the histogram, which has the advantage of 
avoiding misinterpretation of the statistics caused by the granularity of bin 
size used in the histograms, though at the expense of being a less-familiar 
distribution), of the minor- and major-axis residuals are presented in Figure 
\ref{fig::residPlots}, along with the theoretical distributions one would 
obtain from unit-deviation distribution (a Gaussian and an error function, 
respectively).  The distributions agree quite well with the theoretical 
distributions, though there might be a slight excess number in the distribution 
wings.  However, any excess is small, and Gaussian statistics are a reasonable 
approximation for orbit-fitting and companion search applications.

\section{Summary of PHASES Measurements}

The 1332 PHASES measurements are presented in Table 
\ref{phasesAstrometryData}.  All have been processed with the standard pipeline 
described in Section \ref{sec::reduction}.  In total, 51 binaries were 
observed on 
443 nights.  An average of 1983 scans was used for each measurement, and the 
average formal minor-axis uncertainty is 13 $\microas$.  After the scaling 
laws from \S \ref{errorBarCorrection} were applied, the average minor- and 
major-axis uncertainties were 53 $\mu$as and 486 $\mu$as, respectively.

\section{Supporting Measurements}

Infrared differential photometry between the primary and secondary components 
of several PHASES targets were obtained using Keck Adaptive Optics on MJD 
53227.  The differential photometries were measured in the ${\rm K_p}$ filter 
for faint systems, and a narrow band ${\rm H_2}$ 2-1 filter centered at 2.2622 
${\rm \mu m}$ for brighter binaries.  Six systems of moderate brightness were 
measured with both filters.  Differential magnitudes for the 20 binaries 
observed with Keck AO are presented in Table 
\ref{AOPhotometryData}.

\begin{deluxetable}{lllll}
\tablecolumns{5}
\tablewidth{0pc} 
\tablecaption{Keck AO Differential Photometry Measurements \label{AOPhotometryData}}
\tablehead{ 
\colhead{HD} & \colhead{HJD-2453227.5} & \colhead{Filter} & \colhead{$\Delta {\rm m}$} & \colhead{$\sigma$}}
\startdata
5286   & 0.49 & ${\rm H_2}$ & 0.442 & 0.002 \\
6811   & 0.49 & ${\rm H_2}$ & 1.426 & 0.002 \\
13872  & 0.50 & ${\rm H_2}$ & 0.137 & 0.002 \\
17904  & 0.51 & ${\rm K_p}$ & 1.218 & 0.003 \\
17904  & 0.54 & ${\rm H_2}$ & 1.206 & 0.003 \\
40932  & 0.63 & ${\rm H_2}$ & 0.002 & 0.005 \\
41116  & 0.64 & ${\rm H_2}$ & 2.043 & 0.008 \\
44926  & 0.64 & ${\rm H_2}$ & 0.215 & 0.004 \\
137107 & 0.33 & ${\rm K_p}$ & 0.185 & 0.001 \\
140436 & 0.35 & ${\rm K_p}$ & 0.936 & 0.001 \\
165908 & 0.40 & ${\rm H_2}$ & 2.063 & 0.002 \\
165908 & 0.40 & ${\rm K_p}$ & 1.909 & 0.001 \\
171745 & 0.40 & ${\rm K_p}$ & 0.278 & 0.001 \\
171745 & 0.41 & ${\rm H_2}$ & 0.272 & 0.001 \\
171779 & 0.42 & ${\rm K_p}$ & 0.167 & 0.002 \\
171779 & 0.42 & ${\rm H_2}$ & 0.163 & 0.003 \\
176051 & 0.43 & ${\rm H_2}$ & 1.308 & 0.001 \\
176051 & 0.43 & ${\rm K_p}$ & 1.235 & 0.001 \\
196524 & 0.46 & ${\rm H_2}$ & 1.096 & 0.002 \\
196867 & 0.46 & ${\rm H_2}$ & 1.587 & 0.004 \\
202275 & 0.47 & ${\rm H_2}$ & 0.066 & 0.004 \\
206901 & 0.44 & ${\rm H_2}$ & 0.188 & 0.001 \\
207652 & 0.48 & ${\rm K_p}$ & 0.740 & 0.004 \\
213973 & 0.44 & ${\rm K_p}$ & 0.293 & 0.003 \\
221673 & 0.45 & ${\rm H_2}$ & 0.570 & 0.003 
\enddata
\tablecomments{
Column 1 is the star's HD catalog number.  
Column 2 is the heliocentric Julian date offset by the day of observations.  
Column 3 is the filter used during the observation (${\rm K_p}$ or 
${\rm H_2}$).  Columns 5 and 6 are the differential photometry and uncertainty 
measurements in units of magnitudes.
}
\end{deluxetable}

\acknowledgements 
Though it has no impact on the current study, as an astrometry team 
we could not resist pointing out that--due to precession--PHASES target 
HD 149630 ($\sigma$ Her) was the pole star around 8250 BC.
PHASES benefits from the efforts of the PTI collaboration members who have 
each contributed to the development of an extremely reliable observational 
instrument.  Without this outstanding engineering effort to produce a solid 
foundation, advanced phase-referencing techniques would not have been 
possible.  We thank PTI's night assistant Kevin Rykoski for his efforts to 
maintain PTI in excellent condition and operating PTI in phase-referencing 
mode every week.  Part of the work described in this paper was performed at 
the Jet Propulsion Laboratory under contract with the National Aeronautics 
and Space Administration.  Interferometer data were obtained at the Palomar
Observatory with the NASA Palomar Testbed Interferometer, supported
by NASA contracts to the Jet Propulsion Laboratory.  This publication makes 
use of data products from the Two Micron All Sky Survey, which is a joint 
project of the University of Massachusetts and the Infrared Processing and 
Analysis Center/California Institute of Technology, funded by the National 
Aeronautics and Space Administration and the National Science Foundation.  
This research has made use of the Simbad database, operated at CDS, 
Strasbourg, France.  M.W.M. acknowledges support from the Townes Fellowship 
Program, Tennessee State University, and the state of Tennessee through its 
Centers of Excellence program.  Some of the software used for analysis was 
developed as part of the SIM Double Blind Test with support from NASA 
contract NAS7-03001 (JPL 1336910).  
PHASES is funded in part by the California Institute of Technology 
Astronomy Department, and by the National Aeronautics and Space Administration 
under Grant No.~NNG05GJ58G issued through the Terrestrial Planet Finder 
Foundation Science Program.  This work was supported in part by the National 
Science Foundation through grants AST 0300096, AST 0507590, and AST 0505366.
MK is supported by the Foundation for Polish Science through a FOCUS 
grant and fellowship, by the Polish Ministry of Science and Higher 
Education through grant N203 3020 35.

{\it Facilities:} \facility{PO:PTI, Keck I}

\bibliography{main}
\bibliographystyle{apj}

\clearpage
\LongTables

\begin{deluxetable*}{llrrrrrrrrrrrrrrrrrr}
\tablecolumns{20}
\tablecaption{PHASES Astrometry Measurements \label{phasesAstrometryData}}
\tablehead{ 
\colhead{HD} & \colhead{HJD-} & \colhead{$\delta$RA} 
& \colhead{$\delta$Decl}  & \colhead{$\sigma_{\rm min}$} 
& \colhead{$\sigma_{\rm maj}$} & \colhead{$\phi_{\rm e}$} 
&  \colhead{$\sigma_{\rm RA}$} & \colhead{$\sigma_{\rm Dec}$} 
& \colhead{$\frac{\sigma_{\rm RA, Decl}^2}{\sigma_{\rm RA}\sigma_{\rm Decl}}$} 
& \colhead{$\sigma_{\rm min, c}$} & \colhead{$\sigma_{\rm maj, c}$} 
&  \colhead{$\sigma_{\rm RA, c}$} & \colhead{$\sigma_{\rm Decl, c}$} 
& \colhead{$\frac{\sigma_{\rm RA, Decl, c}^2}{\sigma_{\rm RA, c}\sigma_{\rm Decl, c}}$} 
& \colhead{N} & \colhead{LDC} & \colhead{Align} & \colhead{Rate} & 
\colhead{Outlier} \\
\colhead{Number}     & \colhead{2400000.5}             &  \colhead{(mas)}        
& \colhead{(mas)}        & \colhead{($\microas$)}       
& \colhead{($\microas$)}       & \colhead{(deg)}          
& \colhead{($\microas$)}       & \colhead{($\microas$)}     
& \colhead{} & \colhead{($\microas$)}       
& \colhead{($\microas$)}       
& \colhead{($\microas$)}       & \colhead{($\microas$)}     
& \colhead{} & \colhead{}  & \colhead{}    & \colhead{}      &  \colhead{(Hz)} 
&  \colhead{} }
\startdata
4934 & 54004.36328 & -57.0674 & -179.9978 & 27.8 & 416.4 & 161.61 & 395.3 & 134.0 & -0.97578 & 44.7 & 559.1 & 530.8 & 181.4 & -0.96573 & 345 & 1 & 1 & 50 & 0 \\
5286 & 52894.31345 & -681.5589 & 665.6299 & 17.8 & 1360.0 & 146.08 & 1128.5 & 759.1 & -0.99960 & 76.2 & 1773.5 & 1472.3 & 991.7 & -0.99571 & 648 & 0 & 0 & 100 & 0 \\
5286 & 52896.29389 & -681.5838 & 665.5458 & 19.6 & 549.6 & 144.35 & 446.7 & 320.8 & -0.99717 & 82.3 & 728.1 & 593.6 & 429.6 & -0.97200 & 241 & 0 & 0 & 100 & 0 \\
5286 & 52897.28803 & -681.9351 & 665.8532 & 16.4 & 382.3 & 144.32 & 310.7 & 223.4 & -0.99590 & 71.5 & 516.3 & 421.5 & 306.7 & -0.95838 & 566 & 0 & 0 & 100 & 0 \\
5286 & 53198.45485 & -676.2609 & 690.8888 & 3.9 & 213.7 & 143.53 & 171.9 & 127.1 & -0.99928 & 38.0 & 311.1 & 251.2 & 187.4 & -0.96795 & 3129 & 0 & 0 & 100 & 0 \\
5286 & 53215.43340 & -675.9374 & 692.1281 & 7.2 & 273.5 & 146.42 & 227.9 & 151.4 & -0.99838 & 44.4 & 382.1 & 319.3 & 214.6 & -0.96882 & 2038 & 0 & 0 & 100 & 0 \\
5286 & 53222.39706 & -676.3939 & 693.0688 & 4.1 & 217.5 & 144.05 & 176.1 & 127.7 & -0.99921 & 38.3 & 315.5 & 256.4 & 187.8 & -0.96798 & 2583 & 0 & 0 & 100 & 0 \\
5286 & 53229.40966 & -674.7178 & 692.6724 & 17.9 & 1669.5 & 148.35 & 1421.2 & 876.3 & -0.99971 & 76.5 & 2174.9 & 1851.8 & 1143.1 & -0.99691 & 400 & 0 & 0 & 100 & 0 \\
\enddata
\end{deluxetable*}
\clearpage

\end{document}